\documentclass[aps,pra,preprint,amsmath,amssymb]{revtex4}
\usepackage{graphicx,epsfig,epsf}
\usepackage{dcolumn}
\usepackage{bm,psfrag,color}
\usepackage[matrix,frame,arrow]{xypic}
\input{Qcircuit}

\begin{document}
\newcommand{\be}{\begin{equation}}
\newcommand{\ee}{\end{equation}}
\newcommand{\bea}{\begin{eqnarray*}}
\newcommand{\eea}{\end{eqnarray*}}
\newcommand{\ben}{\begin{equation}}
\newcommand{\een}{\end{equation}}
\newcommand{\bean}{\begin{eqnarray}}
\newcommand{\eean}{\end{eqnarray}}

\newcommand{\ri}{{\rm i}}
\newcommand{\re}{{\rm e}}
\newcommand{\bx}{{\bf x}}
\newcommand{\bd}{{\bf d}}
\newcommand{\br}{{\bf r}}
\newcommand{\bk}{{\bf k}}
\newcommand{\bE}{{\bf E}}
\newcommand{\bR}{{\bf R}}
\newcommand{\bM}{{\bf M}}
\newcommand{\bn}{{\bf n}}
\newcommand{\bs}{{\bf s}}
\newcommand{\tbs}{\tilde{\bf s}}
\newcommand{\rSi}{{\rm Si}}
\newcommand{\beps}{\mbox{\boldmath{$\epsilon$}}}
\newcommand{\bthe}{\mbox{\boldmath{$\theta$}}}
\newcommand{\rg}{{\rm g}}
\newcommand{\tr}{{\rm tr}}
\newcommand{\xmax}{x_{\rm max}}
\newcommand{\ra}{{\rm a}}
\newcommand{\rx}{{\rm x}}
\newcommand{\rs}{{\rm s}}
\newcommand{\rP}{{\rm P}}
\newcommand{\up}{\uparrow}
\newcommand{\down}{\downarrow}
\newcommand{\hc}{H_{\rm cond}}
\newcommand{\kb}{k_{\rm B}}
\newcommand{\cI}{{\cal I}}
\newcommand{\tit}{\tilde{t}}
\newcommand{\cE}{{\cal E}}
\newcommand{\cC}{{\cal C}}
\newcommand{\Ubs}{U_{\rm BS}}
\newcommand{\qq}{{\bf ???}}
\newcommand*{\etal}{\textit{et al.}}

\sloppy

\title{Efficiency of Producing Random Unitary Matrices with Quantum
  Circuits}  
\author{Ludovic Arnaud and Daniel Braun}

\affiliation{Laboratoire de Physique Th\'eorique --- IRSAMC, UPS and CNRS,
  Universit\'e de Toulouse, 
  F-31062 Toulouse, FRANCE}  

\begin{abstract}
We study the scaling of the convergence of several statistical properties of
a recently 
introduced random unitary circuit ensemble towards their limits given by
the circular unitary ensemble (CUE). Our study includes the full
distribution of the absolute square of a matrix element, moments of that
distribution up to order eight, as well as correlators containing up to 16
matrix elements in a given column of the unitary matrices. Our numerical
scaling analysis shows that all of these quantities can be reproduced
efficiently, with a number of random gates which scales at most as
$n_q\log (n_q/\epsilon)$ with the
number of qubits $n_q$ for a given fixed precision $\epsilon$. This suggests
that quantities which require an exponentially large number of gates are of
more complex nature.

\end{abstract}
\maketitle

\section{Introduction} 
Random unitary matrices play an important role in many tasks of quantum
information processing, including quantum data hiding \cite{DiVincenzo02},
quantum state distinction  \cite{Sen05}, quantum encryption
\cite{Ambainis04}, superdense coding of quantum states \cite{Harrow04}, and
noise estimation \cite{Emerson05}. In these applications, a random ensemble
of $N\times N$ matrices $U$ drawn uniformly from the Haar measure of the unitary
group, the so--called circular unitary ensemble (CUE), is required
\cite{Mehta91}. In principle, any unitary matrix acting on vectors in the
Hilbert space of dimension $N=2^{n_q}$ of $n_q$ qubits can be approximated
with arbitrary precision using a computationally universal set of quantum
gates that act on one or two qubits at the time
\cite{Deutsch85,DiVincenzo95,Sleator95,Barenco95b,Barenco95a}. However, as 
simple parameter counting quickly confirms, the required number of quantum
gates $n_g$ grows typically exponentially with the number of qubits. Indeed,
${\cal O}(N^2(\ln N)^3)$ gates are required to approximate all matrix
elements of $U$ using a fixed universal gate set \cite{Nielsen00}. This 
makes the 
construction of sets of random unitary matrices which are evenly distributed
according to the Haar measure of the unitary group highly inefficient. One
explicit but inefficient procedure of constructing matrices drawn from CUE is
based on the Hurwitz parametrization (see \cite{Pozniak98}).\\

In a seminal paper, Emerson et
al.~introduced the concept of pseudo-random unitary operators, i.~e.~random
unitary operators which are drawn from a distribution that mimics a uniform
distribution with respect to the Haar measure of the unitary group
\cite{Emerson03}. The 
construction of these operators was motivated by ideas from quantum chaos,
and used a random quantum circuit consisting of random $U(2)$ rotations on
each qubit followed by two qubit-gates that implement an Ising spin
interaction between nearest neighbors. They showed that
this circuit 
produced unitary matrices with a distribution of matrix elements which
converges exponentially with the number of quantum gates to the well--known
distribution of matrix elements of CUE \cite{Haake2000}. Later, Emerson,
Livine, 
and Lloyd 
showed that the joint distribution function of matrix elements of a product
of unitary operators created by a random quantum 
circuit composed of a continuous or discrete universal gate set converges
uniformly and exponentially with the number of quantum gates to the Haar
measure on the unitary group, albeit with a rate which itself decreases
exponentially with the number of qubits \cite{Emerson05}. This left open the
question of the efficiency of the creation of the pseudo-random unitary
operators in the sense of the scaling of the number of gate operations with
the number of qubits. Furthermore, the distribution $P_{ij}(U_{ij})$ of
matrix elements $U_{ij}$ contains only a small amount of information
compared to the full joint distribution of matrix elements. Notably it is
unclear how fast correlators of matrix elements would converge to the CUE
values. 

Quite different statistics have been studied so far for different
random circuit ensembles. In \cite{Znidaric07} the question of the efficient
generation of typical bipartite entanglement between two subsystems was
addressed numerically for a quantum circuit composed of $U(4)$ gates, each
of which was a product of a fixed two--qubit gate and two random single
qubit gates drawn uniformly from the Haar measure of $U(2)$. Exponential
convergence to the CUE value with a rate that depends as $n_q
\ln n_q$ on the number of qubits was found. Oliveira et
al.~introduced the technique of Markov chain analysis to study the same
question and were able to prove an upper bound of ${\cal O}(n_q^3)$ quantum
gates necessary to reach a given (absolute) precision $\epsilon$ for the 
average amount of bipartite entanglement
\cite{Oliveira07,Dahlsten07}. Average gate fidelity was studied in
\cite{Fortunato02}. 
The
distributions of differences between nearest neighbor eigenphases as well as
the distribution of the amount of interference was studied in
\cite{Arnaud07} for the same random unitary circuit ensemble as the one we
will use here (see below). Exponential or even Gaussian convergence was
observed, but the question of the efficiency remained open. 

The study of pseudo-random unitaries is closely related to
the theory of unitary $k$-designs. Dankert et al.~defined a unitary
$k$-design as a discrete set of unitary matrices such that the average of
any polynomial of degree equal or smaller than $k$ in the complex matrix
elements of $U$ over the set equals the average of that polynomial over the
unitary group \cite{Dankert06}. 
Harrow et al.~showed that a random circuit of length polynomial in $n_q$ 
 yields an $\epsilon$--approximate
2--design. Depending on the gate set used, the number of gates $n_g$ needed
to achieve a given precision $\epsilon$ scales as ${\cal O}(n_q(n_q+\log
1/\epsilon))$ or as ${\cal O}(n_q\ln (n_q/\epsilon))$. They also conjectured
that a random circuit on $n_q$ qubits composed of 
poly$(n_q,k)$ random two qubit gates chosen from a universal gate set is an
$\epsilon$--approximate $k$--design \cite{Harrow08}. Originally
unitary designs
were defined for a fixed set of unitary matrices, each of which comes with
the same weight (see \cite{Gross07} for an insightful discussion of their
mathematical 
structure). The definition in \cite{Harrow08} naturally extends the concept
of unitary designs to probability distributions over sets of unitary
matrices, such that each 
random unitary matrix corresponds to a realization of the random quantum
circuit, and averaging over the random circuits realizes the average over
the unitary design. 

The results from \cite{Harrow08} imply that random quantum
circuits can 
efficiently (with ${\cal O}(n_q\ln(n_q/\epsilon))$ gates) reproduce the
CUE averages of $|U_{ij}|^2$,  $|U_{ij}|^4$ and 
$|U_{ij}U_{kl}|^2$. Note that CUE averages of unpaired matrix elements
vanish \cite{Aubert03}. Since entanglement fidelity and gate fidelity can be
expressed as averages over polynomials of order (2,2) in $U_{ij}$ and
$U_{kl}^*$, this result 
confirms the numerical finding in \cite{Znidaric07}.

In this paper we study numerically the efficiency with which the random
unitary circuit ensemble (UCE) introduced in \cite{Arnaud07} (see below for
its 
definition) reproduces various
statistical properties of the matrix elements of CUE matrices. Our study
includes the full distribution of the (absolute square of) matrix elements,
moments of that 
distribution up to order $|U_{ij}|^{16}$, as well as correlators containing
containing up to $16$ matrix elements of  
a given column of the unitary matrix. Within the range of numerically
accessible 
sizes of the quantum circuits (up to 28 qubits for the distribution of a
single matrix element, down to 15 qubits for the correlators), our results
show that, surprisingly, the 
number of gates required to 
reach a given precision $\epsilon$ for all of these quantities grows no
faster than $n_q  
\ln (n_q/\epsilon)$, indicating that the statistical
properties of CUE which require an exponential number of quantum gates in
order to be well approximated by a random quantum circuit,
must be of a more sophisticated nature. 
 
\section{Convergence of UCE to CUE}
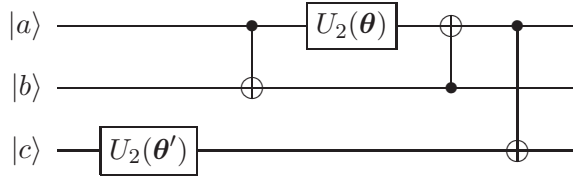
\begin{figure}[h]
\begin{minipage}{1.00\textwidth}
\Qcircuit @C=1.5em @R=1em {
\lstick{\ket{a}} & \qw & \ctrl{1} & \gate{U_{2}(\bthe)} & \targ & \ctrl{2} & \qw \\
\lstick{\ket{b}} & \qw & \targ & \qw & \ctrl{-1} & \qw & \qw \\
\lstick{\ket{c}} & \gate{U_{2}(\bthe')} & \qw & \qw & \qw & \targ & \qw
}
\end{minipage}
\caption{A random UCE circuit. The two different angles $\bthe$ and $\bthe'$ mean two different random U(2) gates.}
\label{fig.circuit} 
\end{figure}

The unitary circuit ensemble (UCE) introduced in \cite{Arnaud07} consists 
of quantum 
algorithms which use two kinds of quantum gates: $U(2)$ gates which act on
single qubits, and the CNOT
gate which acts on two qubits at the time. Each algorithm is built from a
random 
sequence of 
these gates, where the probability that a given gate is a 1-qubit gate is
$p_g$ and 
the probability that it is a 2-qubit gate is $1-p_g$. We set $p_g=0.5$
throughout this 
paper. The choice of the qubit(s) on which a gate acts, is made uniformly
 and independently for different gates over all the qubits. Fig
 (\ref{fig.circuit}) shows an example of this kind 
of algorithm for 3 qubits and 5 gates. 
The $U(2)$ gates are chosen uniformly with respect to the invariant Haar
measure 
of the U(2) group. They can be parametrized with four angles $\alpha$,
$\psi$, $\chi$ chosen randomly and uniformly 
from $[0,2\pi [$, and $\varphi=\arcsin(\xi^{1/2})$ with $\xi$ picked
    randomly and 
uniformly from $[0,1]$,
\be \label{U2}
U_2(\bthe)=
\re^{\ri \alpha}\left(\begin{array}{cc}
  \cos\varphi\re^{\ri\psi}&\sin\varphi\re^{\ri\chi}\\
  -\sin\varphi\re^{-\ri\chi}&\cos\varphi\re^{-\ri\psi}\\
\end{array}
\right)\equiv
\left(\begin{array}{cc}
  c&s\\
  -\bar{s}&\bar{c}\\
\end{array}
\right)\,,
\ee\\
where we have abbreviated $\bthe=(\alpha,\psi,\chi,\varphi)$
\cite{Pozniak98}. The phases 
$\alpha$ only modify the global phase of the algorithm and are irrelevant
for the statistical properties that we are going to study. From the results
of \cite{Emerson05} it is clear that in
the limit of the number of gates $n_g\to\infty$ and fixed $n_q$, UCE converges to CUE.

The UCE gate set might be
summarized as $\Gamma=\{
\{\frac{d\mu(U_2)}{4},U_2\otimes {\bf 1}_2\},
\{\frac{d\mu(U_2)}{4},{\bf 1}_2\otimes U_2\},
\{\frac{1}{4},U_{\rm  CNOT\,1,2}\},
\{\frac{1}{4},U_{\rm  CNOT\,2,1}\}
\}$, where the first number in each pair in the list is the probability
  that
the second member of the pair will be selected in any step of the algorithm,
$\mu(U_2)$ means the Haar measure of $U(2)$, and $U_{\rm  CNOT\,i,j}$ is a
  controlled-NOT gate with control qubit $i$ and target qubit $j$. It is
  easily checked 
that $\Gamma$ is a ``2--copy gapped gate set'' in the terminology of
\cite{Harrow08}. This means that the operator $G=\int_{U(4)}
U\otimes U\otimes U^*\otimes U^*d\mu_\Gamma(U)$, defined for a general gate
set  distributed continuously over $U(4)$ with measure $\mu_\Gamma(U)$,  has
only two   eigenvalues with 
absolute value equal to 1. The difference between this largest degenerate
eigenvalue
1 and the next smaller eigenvalue (in terms of its absolute value) is called
the spectral gap $\Delta$.  Our gate set $\Gamma$ has spectral gap
$\Delta\simeq 0.232703$, if the gates are represented as $4\times 4$
matrices.  The gap is expected to decay as $1/n_q$ if the gates are
represented as matrices of size $2^{n_q}$, but will be finite for
any finite $n_q$ \cite{Harrow08.2}. 

\subsection{Distribution of matrix elements}
The uniform distribution of CUE matrices of size $N$ with respect to the
Haar measure of the unitary group $U(N)$
yields a specific joint probability distribution
$P(U)\equiv P(U_{11},U_{12},...,U_{NN})$ of the matrix elements $U_{ij}$
which 
entirely defines this ensemble. Convergence of UCE to CUE means that the
joint probability distribution $\tilde{P}(U)$ associated with UCE converges
to $P(U)$. However, a direct numerical study of the joint distribution is
impractical since the number of necessary realizations grows
exponentially 
with the number of independent arguments of $\tilde{P}(U)$. More practical
quantities can be obtained from the joint probability distribution by
integrating out several variables. A natural quantity to consider is the
distribution of matrix elements, which depends on one complex variable, and
which is 
obtained by integrating out the other $N^2-1$ complex parameters, 
\be
P(U_{ij})=\int...\int \prod_{(k,l) \ne (i,j)}dU_{kl} P(U).
\ee

The first quantity we study in this paper, closely related to $P(U_{ij})$,
is the distribution of quantities defined by $l_{ij}=\ln(N|U_{ij}|^2)$. For
CUE, one shows in random matrix theory (RMT) that all $l_{ij}$ are
distributed according to the normalized distribution 
\be
P(l)=\frac{(N-1)}{N}e^{l}\left(1-\frac{e^{l}}{N}\right)^{N-2},
\ee
independently of the choice of the index of $l_{ij}$ \cite{Haake2000}.\\

For UCE, the distribution of matrix elements is not independent of the
elements chosen as long as the number of gates $n_g$ is small, but becomes
uniform over the matrix in the limit $n_g\to\infty$. For numerical
efficiency, 
we have made two simplifications:\\
First, we produce and propagate only the first column of the matrix. This
obviously reduces drastically the memory requirement, and moreover, the
action of a CNOT gate on this vector requires only the manipulation of a
subset of the matrix elements. With the binary notation of the row index $i$
of a matrix element $U_{i1}$,
$i=1+\sum_{\alpha=1}^{n_q}\sigma_{\alpha}2^{\alpha}$, a 
CNOT between qubits $k$ (control) and $l$ (target) in $[1,n_q]$ requires
only the exchange of the $2^{n_{q}-2}$ elements in positions where
$(\sigma_{k}=0,\sigma_{l}=1)$  with the $2^{n_{q}-2}$ elements in positions
where $(\sigma_{k}=1,\sigma_{l}=1)$. For the $U(2)$ gates, each element in
the new 
column is a linear combination of two old elements, with $c,s,-\bar{s}$, or
$\bar{c}$ as coefficients.

Secondly, we define $\tilde{P}(l)$ by averaging
both over the realizations ($\langle\ldots\rangle_{R}$) and the
elements in the $1^{st}$ column  ($\langle\ldots\rangle_{C}$) , 
\ben \label{PUCE}
\tilde{P}(l)=\frac{1}{n_{r}N}\sum_{r=1}^{n_r}\sum_{i=1}^{N}\tilde{h}\left(l_{i1}^{(r)}\right)\equiv\langle\,\langle \tilde{h}(l) \rangle_{C}\rangle_{R}
\een
where $\tilde{h}(l_{i1}^{(r)})$ is the histogram for the $i^{th}$ component in the
first column of the $r^{th}$ matrix. In order to obtain good statistics, it
is important to produce a large enough number of matrices $n_{r}$. However,
for a given number of matrices, when one adds one qubit, the calculation
time roughly doubles, because the size of the Hilbert space is doubled. For
this reason, as long as $n_q\le 20$, we considered an ensemble of
$n_{r}=a\,2^{b-n_{q}}$  
matrices (with a and b integers). In
other words, when adding a qubit, the increase of the size of the Hilbert
space by a factor 2 is compensated by reducing the size of the ensemble by
a factor 2, without a loss of statistics. We choose $a=10$ and $b=20$
leading to a total of about $10^7$ matrix elements. Due to the correlations
between the matrix elements  (see below), averaging over a column is not
quite as effective as averaging over realizations, and therefore the noise
of the numerical data increases with growing $n_q$. For more than 20 qubits,
numerical run time limitations forced us to fix the number of realizations to 
$n_{r}=10$.

\begin{figure}[h]
\epsfig{file=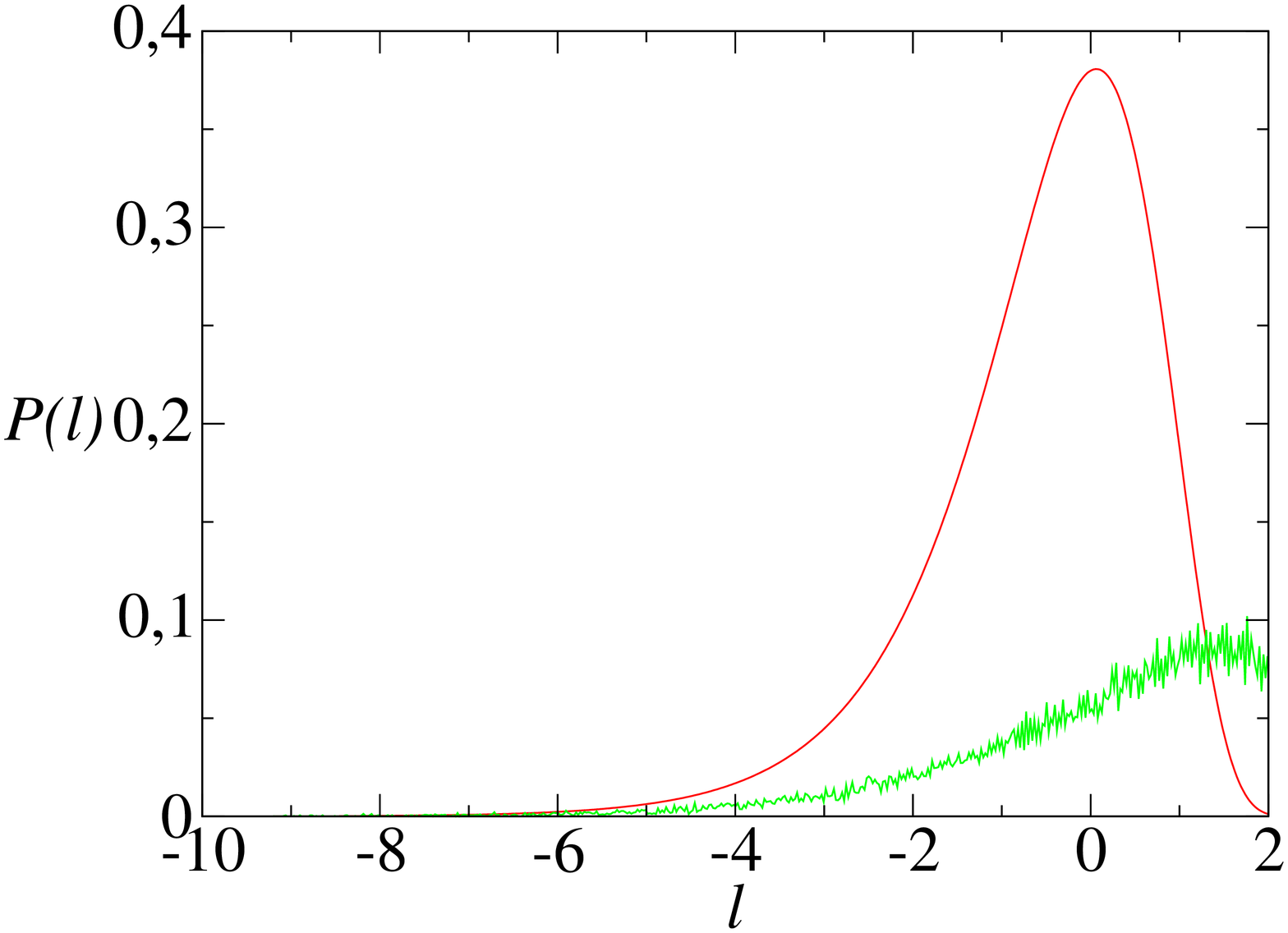,width=4.5cm,angle=270}
\epsfig{file=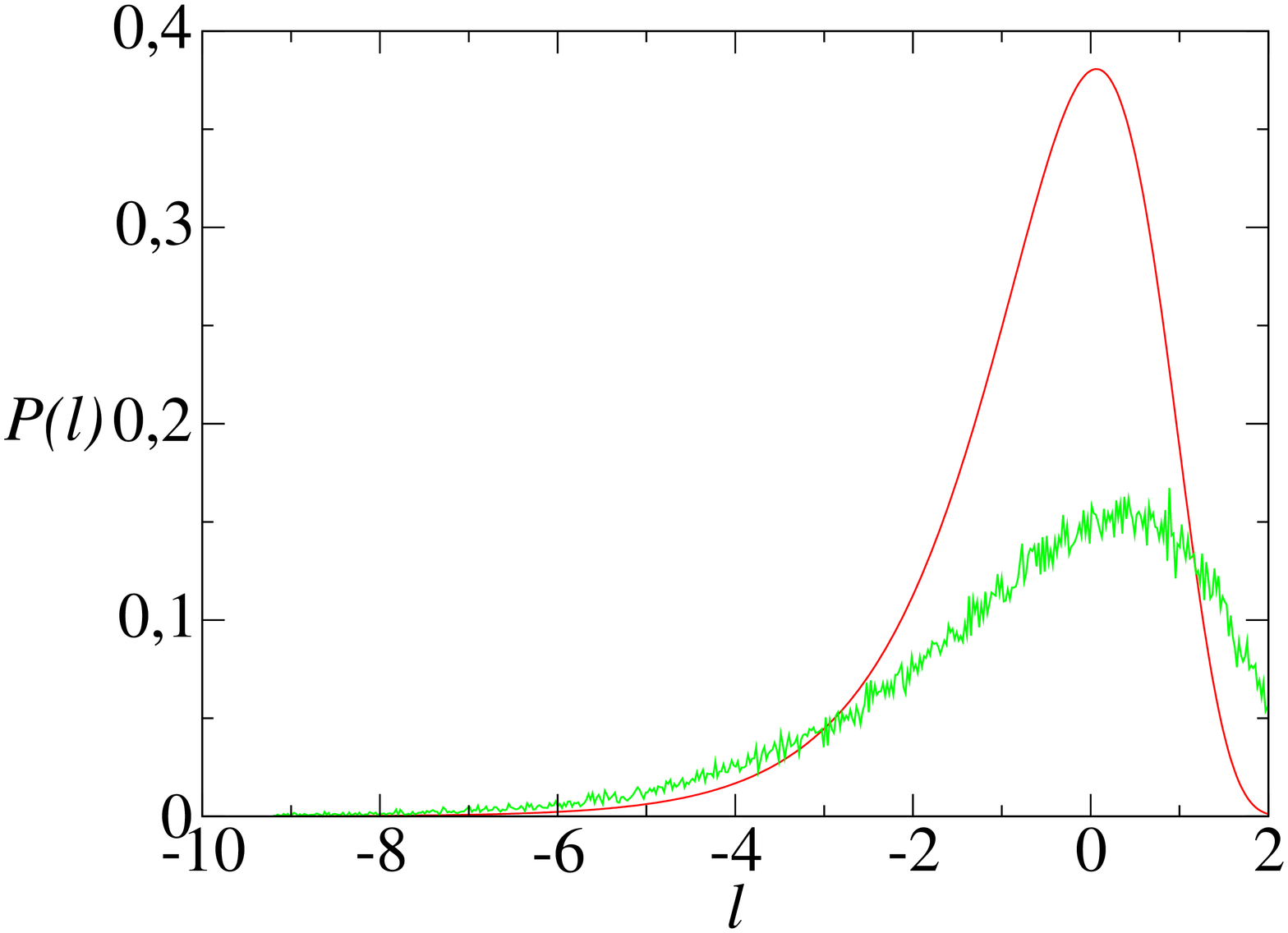,width=4.5cm,angle=270}\\
\epsfig{file=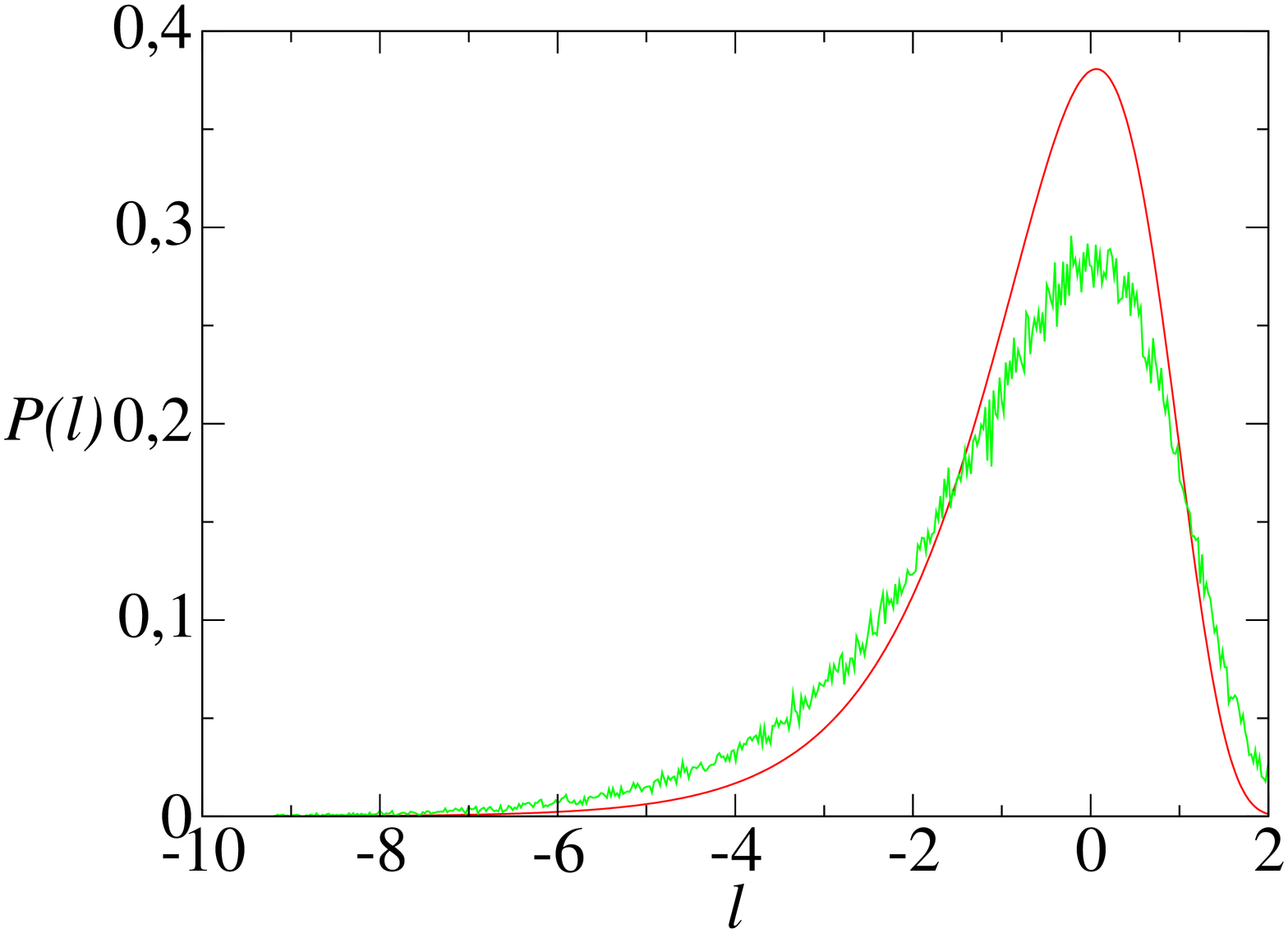,width=4.5cm,angle=270}
\epsfig{file=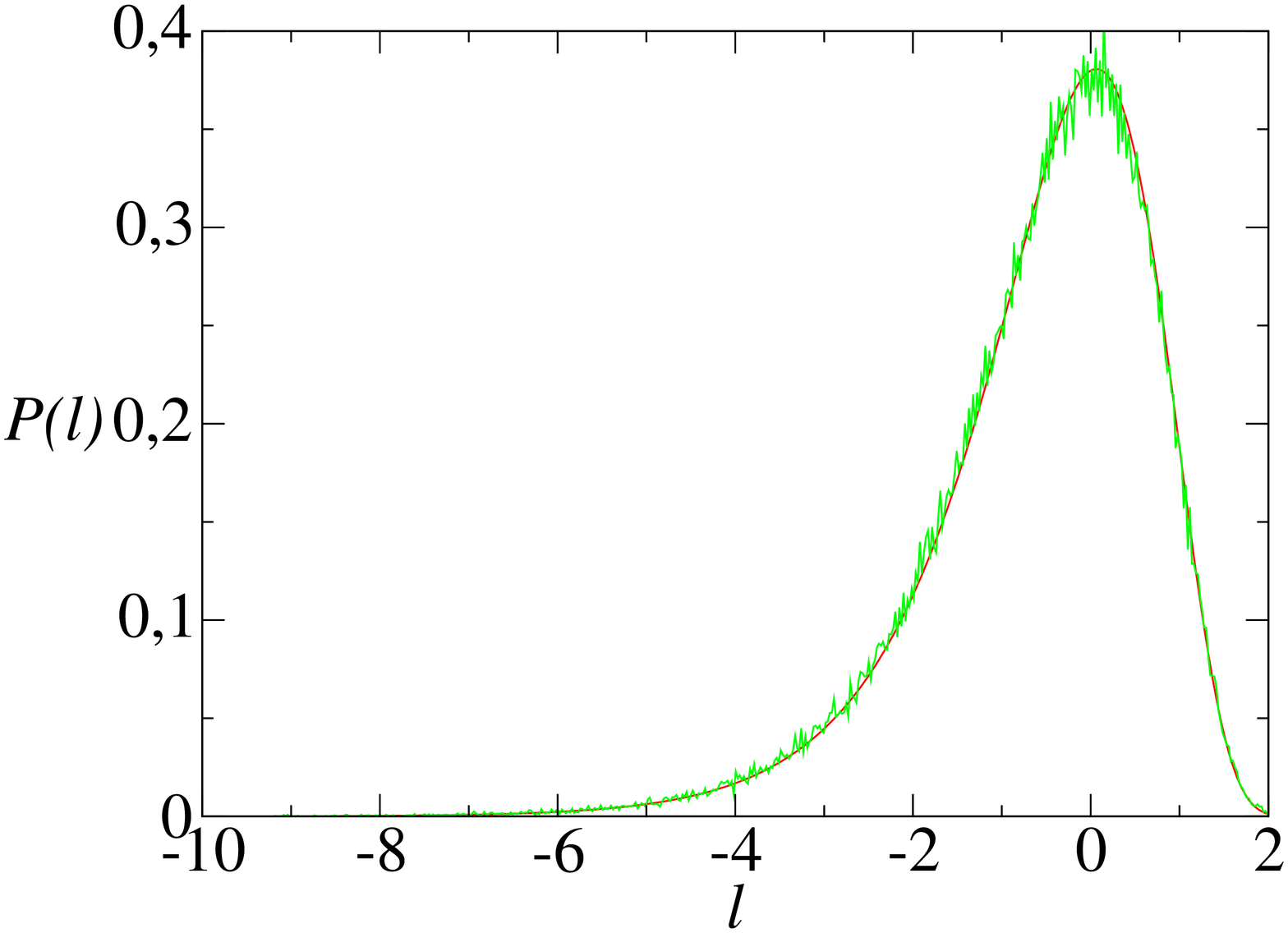,width=4.5cm,angle=270}
\caption{Convergence of $\tilde{P}(l)$ to $P(l)$ (dashed line) for 4 qubits with $n_g=$5,
  10, 20 and 50 for an ensemble of $10^4$ matrices}
\label{fig.disty}        
\end{figure}
\begin{figure}[h]
\epsfig{file=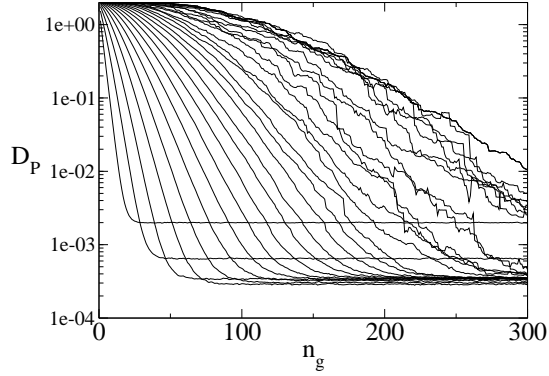,width=6cm,angle=270}
\caption{The distance $D_P(n_{g})$ between the distributions $P(l)$ and
  $\tilde{P}(l)$ as function of the number of gates $n_g$ for
  $n_q=2,3,\ldots,28$ qubits (from left to right).} 
\label{fig.distfid}        
\end{figure}

\begin{figure}[h]
\epsfig{file=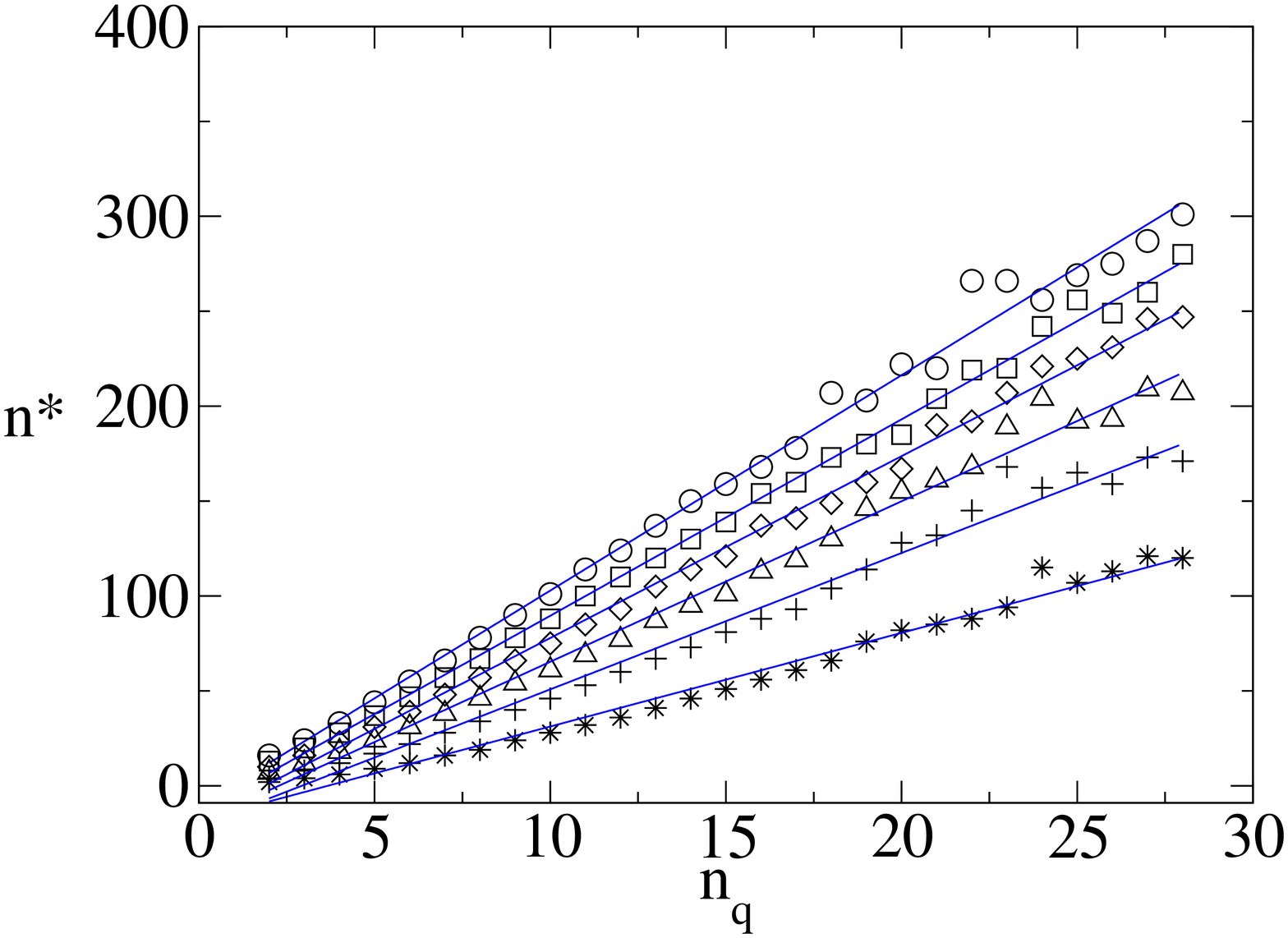,width=4.5cm,angle=270}
\epsfig{file=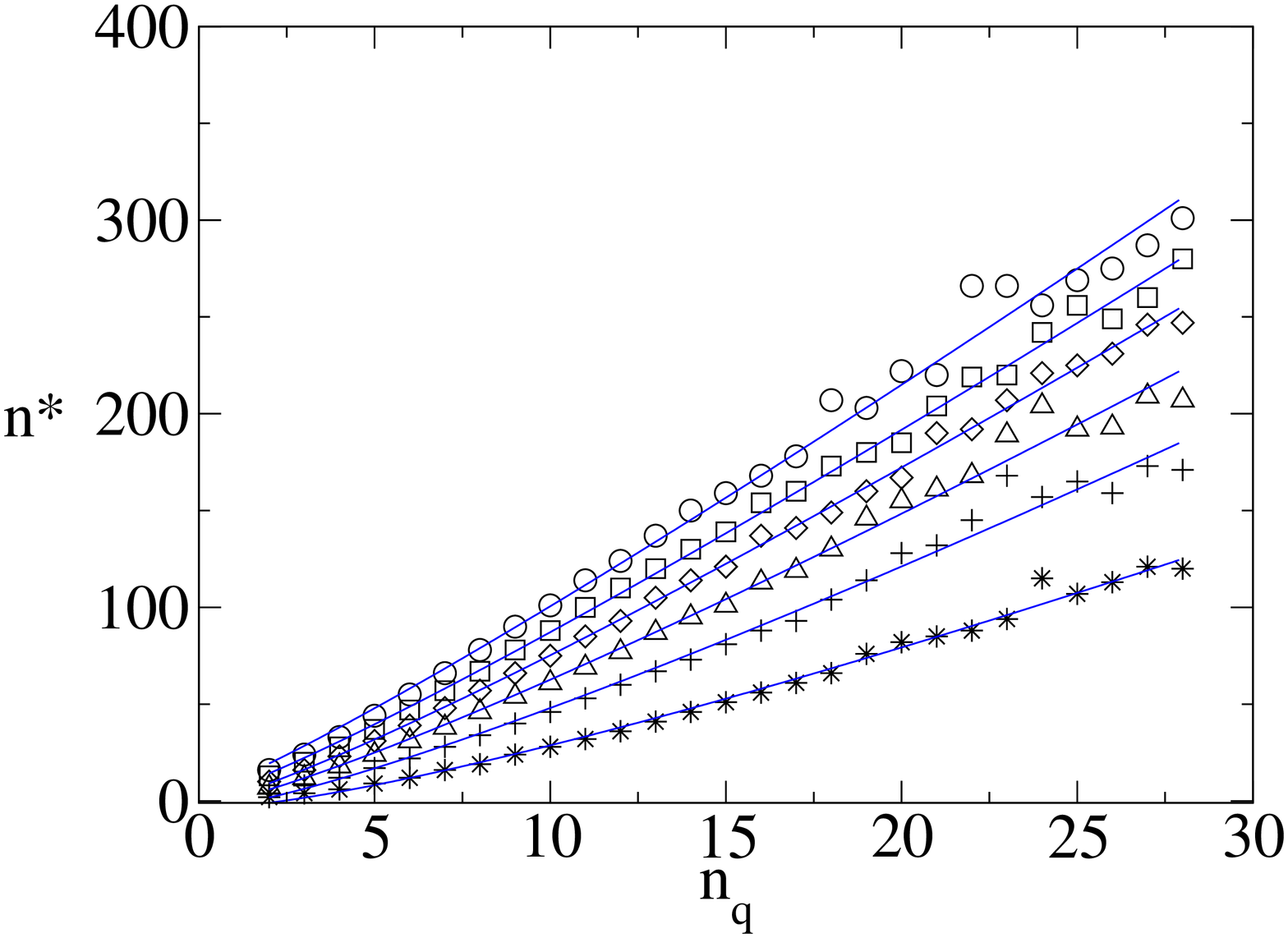,width=4.5cm,angle=270}\\
\epsfig{file=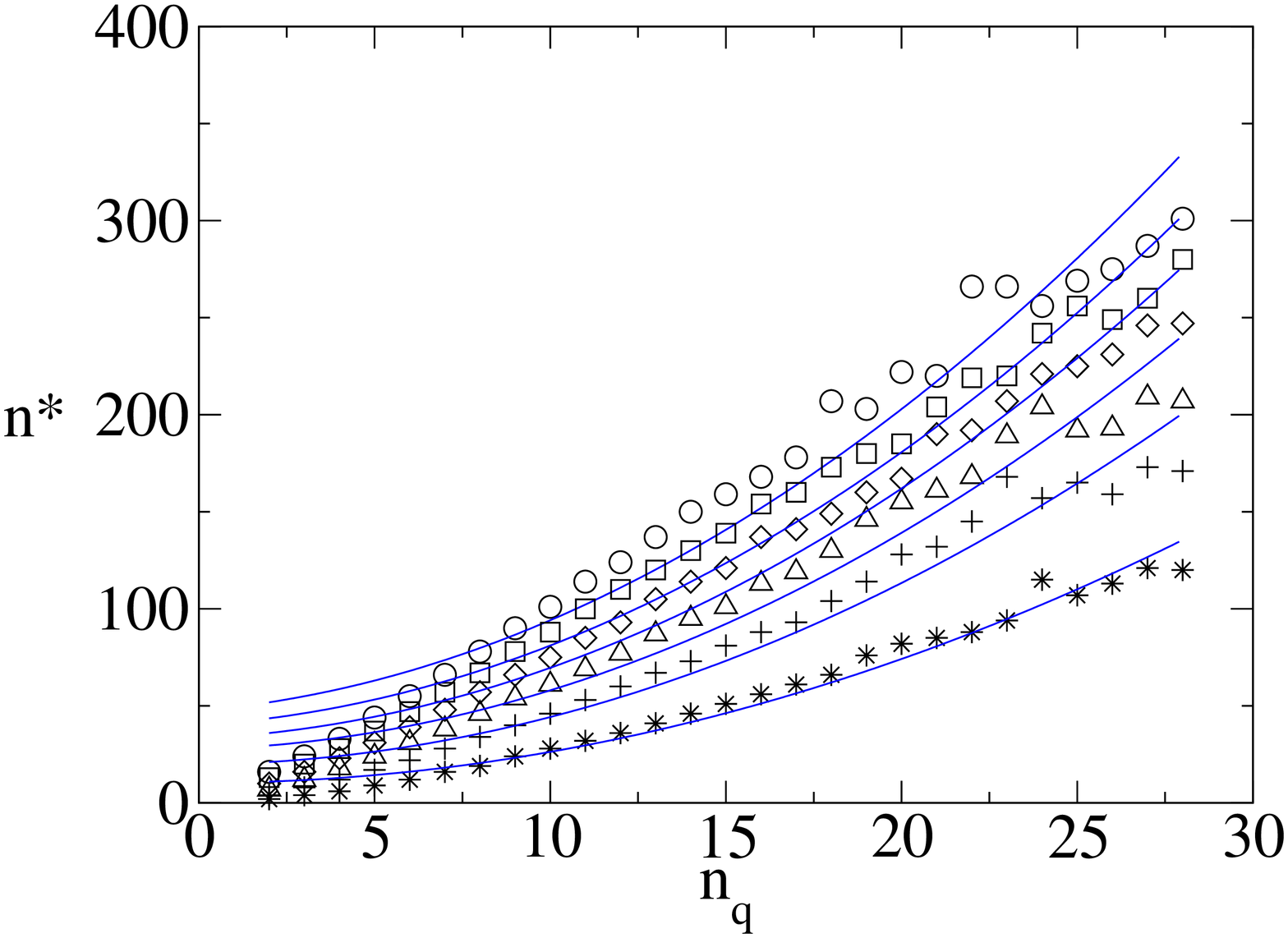,width=4.5cm,angle=270}
\epsfig{file=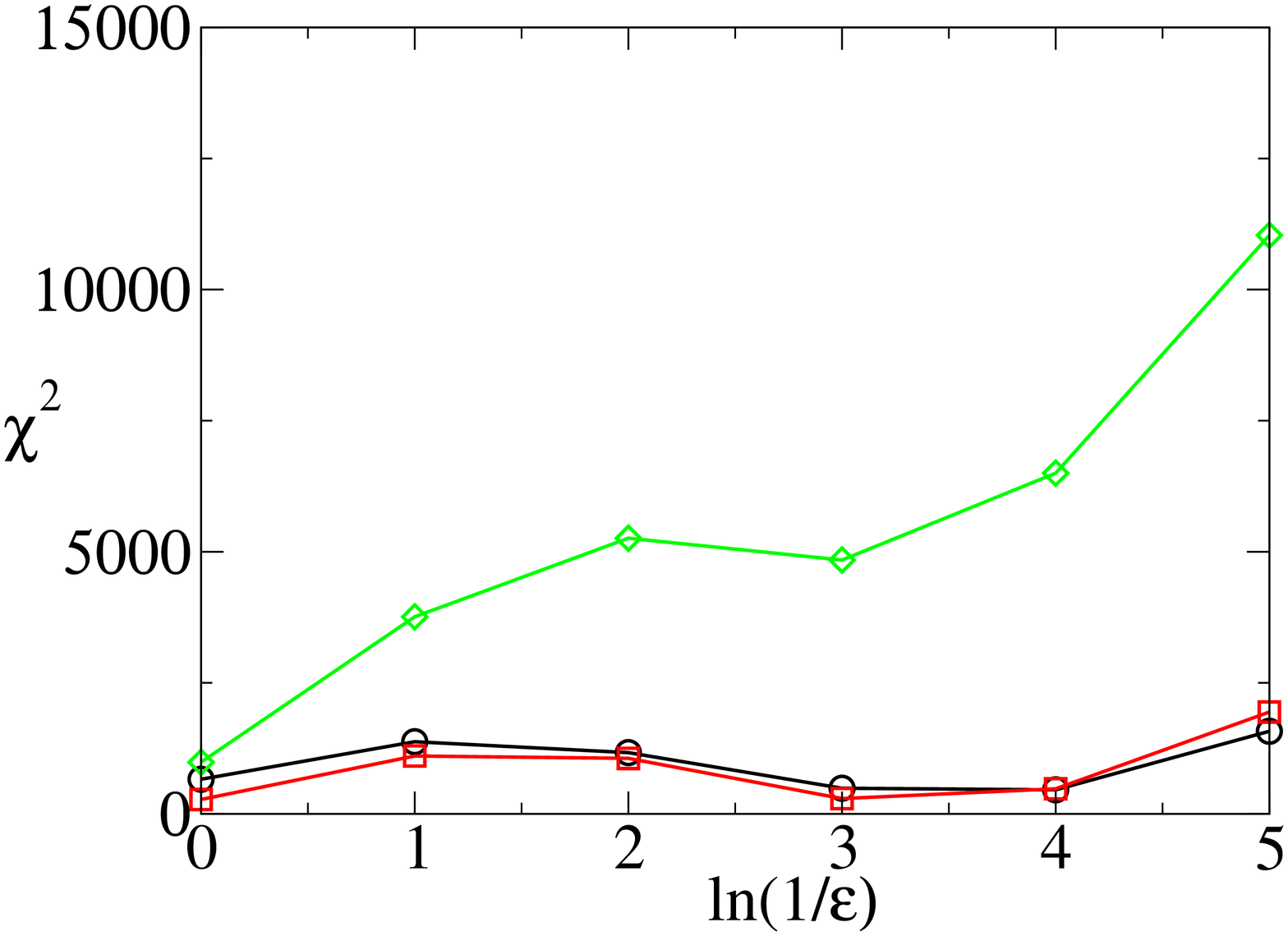,width=4.5cm,angle=270}
\caption{(Color online) The number of gates $n^*$ needed to achieve $D_p\le
  \epsilon$ for $\ln(\epsilon)=$0,-1,-2,-3,-4 and -5 ($\ast$, $+$, $\triangle$, $\diamond$, $\Box$, $\circ$  respectively) and
  $n_q=2\ldots 28$. Straight lines are fits to the functions $f_1$, $f_2$ and
  $f_3$ (1st, 2nd and 3rd plot respectively). The last plot shows $\chi^2$
  for these fits ($f_1$ ($\circ$), $f_2$ ($\Box$)), and $f_3$ ($\diamond$).}  
\label{fig.distfit}        
\end{figure}  

\begin{figure}[h]
\epsfig{file=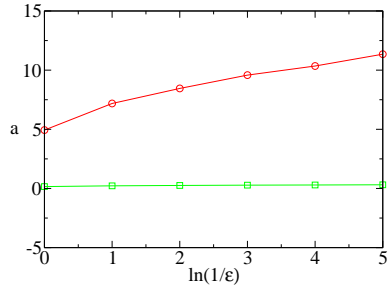,width=4.5cm,angle=270}
\caption{(Color online) The coefficients $a_1$ (green squares) and $a_2$
  (red circles) as a
  function of $\ln(1/\epsilon)$. } 
\label{fig.distpentelin}        
\end{figure}

Fig.~\ref{fig.disty} shows, for $n_q=4$,  
the convergence of $\tilde{P}(l)$ to $P(l)$  with increasing $n_g$.
To quantify the scaling of the convergence with the number of qubits, we define
the quantity 
\bean \label{distance_p}
D_P&=&\int_0^\infty 
\left(\sqrt{\tilde{P}(l)}-\sqrt{P(l)}\right)^2\,dl \nonumber\\
&=&2\left(1-\int_0^\infty\sqrt{\tilde{P}(l)P(l)}\,dl\right) \le 2\nonumber
\eean
which represents a distance between the square roots of the UCE and CUE
distributions. This distance goes to zero as UCE converges toward CUE for
$n_g\to\infty$. 
Using the square roots rather than the distributions themselves is motivated
by the fact that $D_p$ is bounded from above by the value two, which
simplifies the scaling analysis. 
Fig.~\ref{fig.distfid} shows the behavior of $D_P$ as function of $n_{g}$ for
$n_q=2,3,\ldots,28$. As expected, this quantity decays rapidly when 
the number of gates $n_{g}$ grows, but the decay slows down with increasing
$n_q$. Since our numerical calculations use 
a finite number of realizations,  $\tilde{P}(l)$ fluctuates about
$P(l)$. The distance $D_P$ can therefore never vanish exactly, and we observe
that it 
saturates for large $n_g$ at a finite level $d_{min}$ which depends on
$n_q$. The level of 
saturation 
can be reduced by increasing $n_r$. When $D_P$ saturates, our
ensemble becomes indistinguishable from CUE within the precision of the
numerics. We have fitted 
$D_P$ for each value of $n_q$ and we observed that when $n_q$ is small
($n_q\lesssim 12$), $D_P$ is well fitted by $2\,e^{-\alpha\,n_q}$ whereas
for larger 
values of $n_q$, $D_P$ has a pronounced quadratic component in its exponent
($D_P\simeq2\,e^{-\alpha\,n_q-\beta n_q^2}$). The change in the functional 
dependence of $D_P$ on $n_q$ makes it difficult to determine the scaling of
the rate of convergence with $n_q$. We have therefore preferred to base our
analysis on the number of gates $n^*$ needed to achieve a fixed small value
$\epsilon$ of $D_P$ for a given number of qubits. Figure \ref{fig.distfit}
shows the 
behavior of $n^*$ as function of  $n_q$ for six different values of
$\epsilon$ ($\ln(\epsilon)$ between -5 and 0). We have fitted $n^{*}(n_q)$
with three different 2-parameters functions, 
\bean \label{fit_function}
f_1&=& a_1\,n_{q}+b_1\,, \\
f_2&=& a_2\,n_q\ln(n_q/\epsilon)+b_2\,,\\ 
f_3&=& a_3\,n_q(n_q+\ln(1/\epsilon))+b_3\,,
\eean

The linear function $f_1$ is an obvious choice given the appearance of the
numerical data. The functions $f_2$ and $f_3$ are motivated by the results  
in \cite{Harrow08} on 2--designs. These authors defined the convergence of
unitary $k$--designs by the action on a test density matrix $\rho$ of
dimension $k 2^{n_q}$. As measure of distance, they consider the completely
bounded (``diamond'')
norm of the difference between the state ${\cal
  G}_W(\rho)=\sum_ip_iU_i^{\otimes k}\rho(U_i^\dagger)^{\otimes k}$
propagated by the $k$--design and ${\cal
  G}_H(\rho)=\int_U U^{\otimes k}\rho(U^\dagger)^{\otimes k}$ resulting from
the propagator averaged over the unitary group. The gate set
$\Gamma=\{\{p_i, U_i\}\}$ 
  of unitary matrices 
$U_i$ together with their probabilities $p_i$ need to form 
a 2--copy gapped gate set. They show that a random
quantum circuit of length $n_g$ drawn from a 2--copy gapped gate set is an
$\epsilon$--approximate unitary 2--design if $n_g\ge
C(n_q(n_q+\log(1/\epsilon)))$ with some positive constant $C$ which may
depend on the gate set. In the special case of a gate set drawn uniformly
from $U(4)$, which has maximum spectral gap $\Delta=1$ (i.e.~$G$ is a
projector), it was found that an $\epsilon$--approximate unitary 2--design
is already reached for $n_g\ge Cn_q\log
(n_q/\epsilon))$. 

As mentioned, our gate set $\Gamma$ is indeed two-copy gapped, with spectral
gap $\Delta\simeq 0.232703$, and the results of \cite{Harrow08} do therefore
apply to the convergence of UCE to CUE. While one should be cautious in
directly comparing these results, which constitute an upper bound, and are
based on the  
propagation of a trial state and the use of the diamond norm with our
results which use  $D_p$ as a measure of distance, it seems plausible that
the convergence of the 
distribution of matrix elements of the propagator should be related to the
convergence of a propagated test state (see also \cite{Toth07}, where an
efficient quantum algorithm for twirling was introduced). Based on the above
cited results of 
\cite{Harrow08} the function $f_3$ would therefore be the most natural
candidate for a fit of $n^*(n_q)$. However, it turns out that the function
$f_2$, even though relevant {\em a priori} for spectral gap $\Delta=1$, fits
our 
numerical data much better, i.e.~in what concerns the distribution of matrix
elements, UCE converges to CUE much more rapidly than
expected from the lower bound on $n_g$ mentioned.

The quality of the fits is
measured by $\chi^2$, the sum of squares of deviations (see
Fig.~\ref{fig.distfit}). We see that the simple linear
behavior $f_1$ fits better than the quadratic form despite a slight
upwards curvature of the curves $n^*(n_q)$. That curvature is well captured by
the $n_q\ln n_q$ behavior of $f_2$, whereas the quadratic behavior of $f_3$
fits much worse. The function $f_2$ gives in
addition the correct $\epsilon$--dependence, i.e.~a coefficient
$a_2\simeq 0.2$ which 
is basically independent of $\epsilon$ (see Fig.~\ref{fig.distpentelin}). 

A clear distinction between
$f_1$ and $f_2$ is not possible based on the numerical data, as both fit
very well in 
the limited range of $n_q$ available. Nevertheless, our numerical results
clearly indicate that, concerning the
distribution of matrix elements, CUE 
can be efficiently simulated by UCE, in the sense that the number of gates
used to achieve a given level of accuracy $\epsilon$ grows only like $n_q\ln
(n_q/\epsilon)$ with 
the number of qubits, and in any case more slowly than $n_q^2$. This is rather
surprising, as $\tilde{P}(l)$ contains 
information about all moments, and one is therefore led to the conclusion
that no 
moment of appreciable weight in the reconstruction of the distribution
should need more than ${\cal O}(n_q\ln
(n_q/\epsilon))$ gates before coming within 
distance $\epsilon$ relative to the
CUE value. In order to confirm this hypothesis, we have 
studied several $k$-th moments directly.

\subsection{Moments of the distribution of matrix elements}
\begin{figure}[h]
\epsfig{file=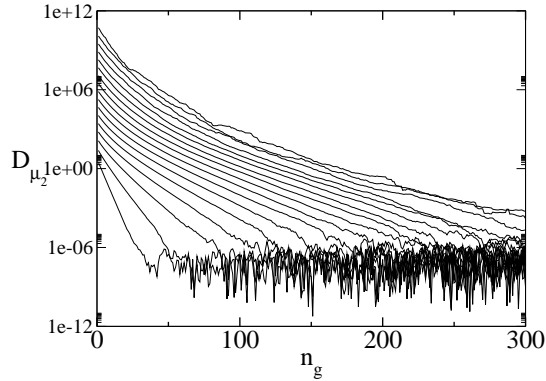,width=6cm,angle=270}
\caption{$D_{\mu_2}(n_{g})$ for a number of qubits $n_q$ varying between $2$
  and $18$ (from left to right).} 
\label{fig.mom2fid}        
\end{figure}

The $k$-th moment $\mu_k$ of the distribution of matrix elements is defined
as $\mu_k=\langle y^{k} \rangle=N^k\langle 
|U_{ij}|^{2\,k} \rangle$. Invariant integration \cite{Aubert03} leads for
CUE to  
\begin{equation} \label{muk}
\mu_k=\frac{k\,!\,N^{k}(N-1)!}{(N+k-1)!}\,,
\end{equation}
which tends to $k\,!$ for $N\to\infty$ and $k$ fixed.
For UCE we average again over both random realizations and elements in the
first column of $U$, analogously to (\ref{PUCE}), and define the $k$-th moment
as 
\ben \label{muUCE}
\tilde{\mu}_k=\frac{1}{n_{r}}\sum_{r=1}^{n_r}\frac{1}{N}\sum_{i=1}^{N}\left(y_{i1}^{(r)}\right)^k=\langle\,\langle y^k \rangle_{C}\rangle_{R}\,.
\een
where $y_{i1}^{(r)}$ is equal to $N|U(l_{i1}^{(r)})|^2$ for the $i^{th}$
component in the first column of the $r^{th}$ matrix. As measure of the deviation from the CUE result we use the relative deviations
\bean \label{distance_mu}
D_{\mu_k}&=&\frac{|\tilde{\mu_k}-\mu_k|}{\mu_k}\,.
\nonumber
\eean
 We have calculated $D_{\mu_k}$ for k=2, 4, and 8. For the latter two cases,
 we used $n_r=10^5$ for 
$n_q=2,\ldots,14$ and $n_r=5\cdot 10^4$ for 15 qubits. 
Fig.~\ref{fig.mom2fid} shows the behavior of 
 $D_{\mu_2}(n_{g})$ for $n_q=2,\ldots,18$. The curves for 
$D_{\mu_4}$ and 
$D_{\mu_8}$ look very similar, with the exception of a higher
saturation 
level, and are not shown.  

\begin{figure}[h]
\epsfig{file=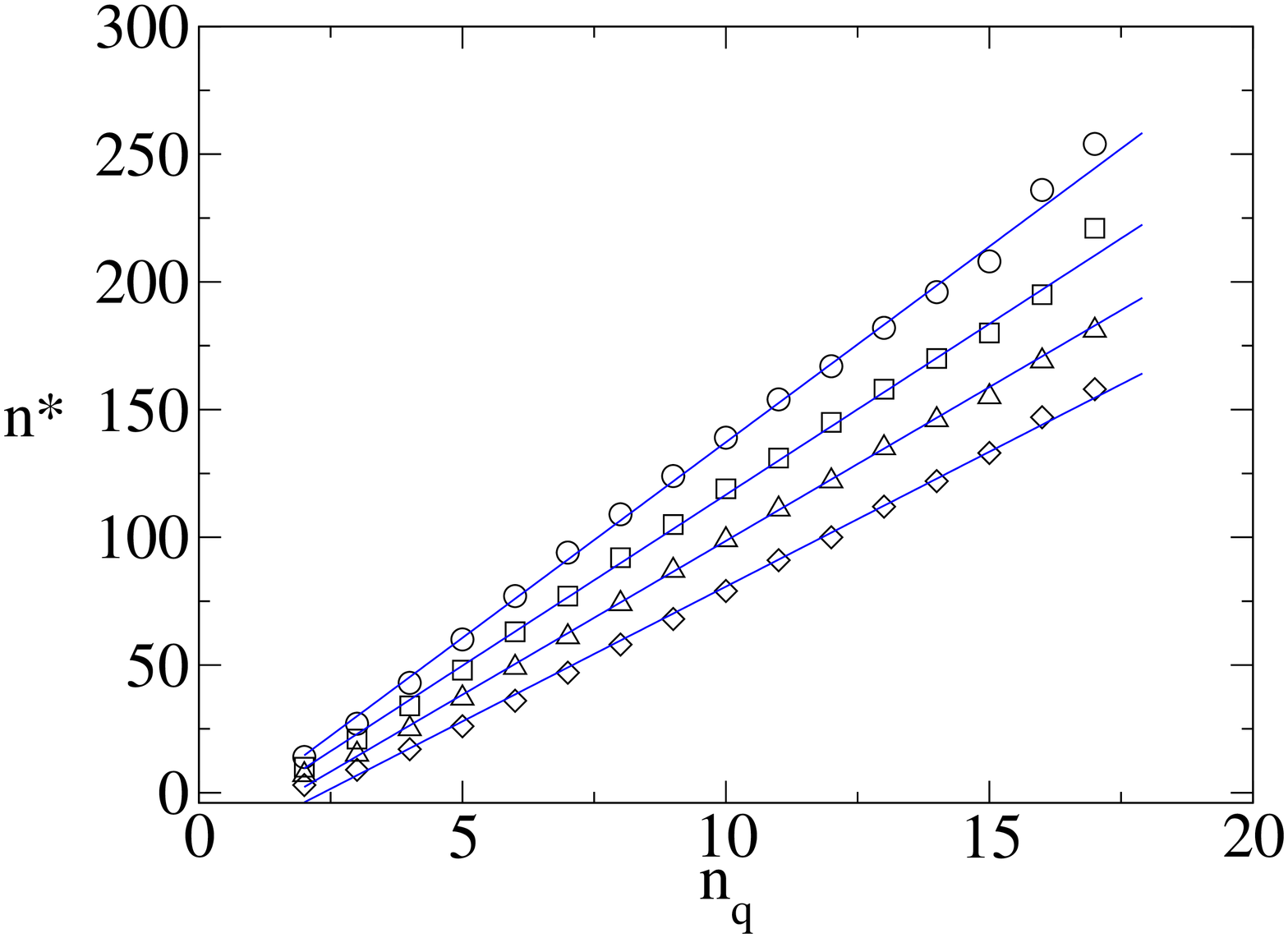,width=4cm,angle=270}
\epsfig{file=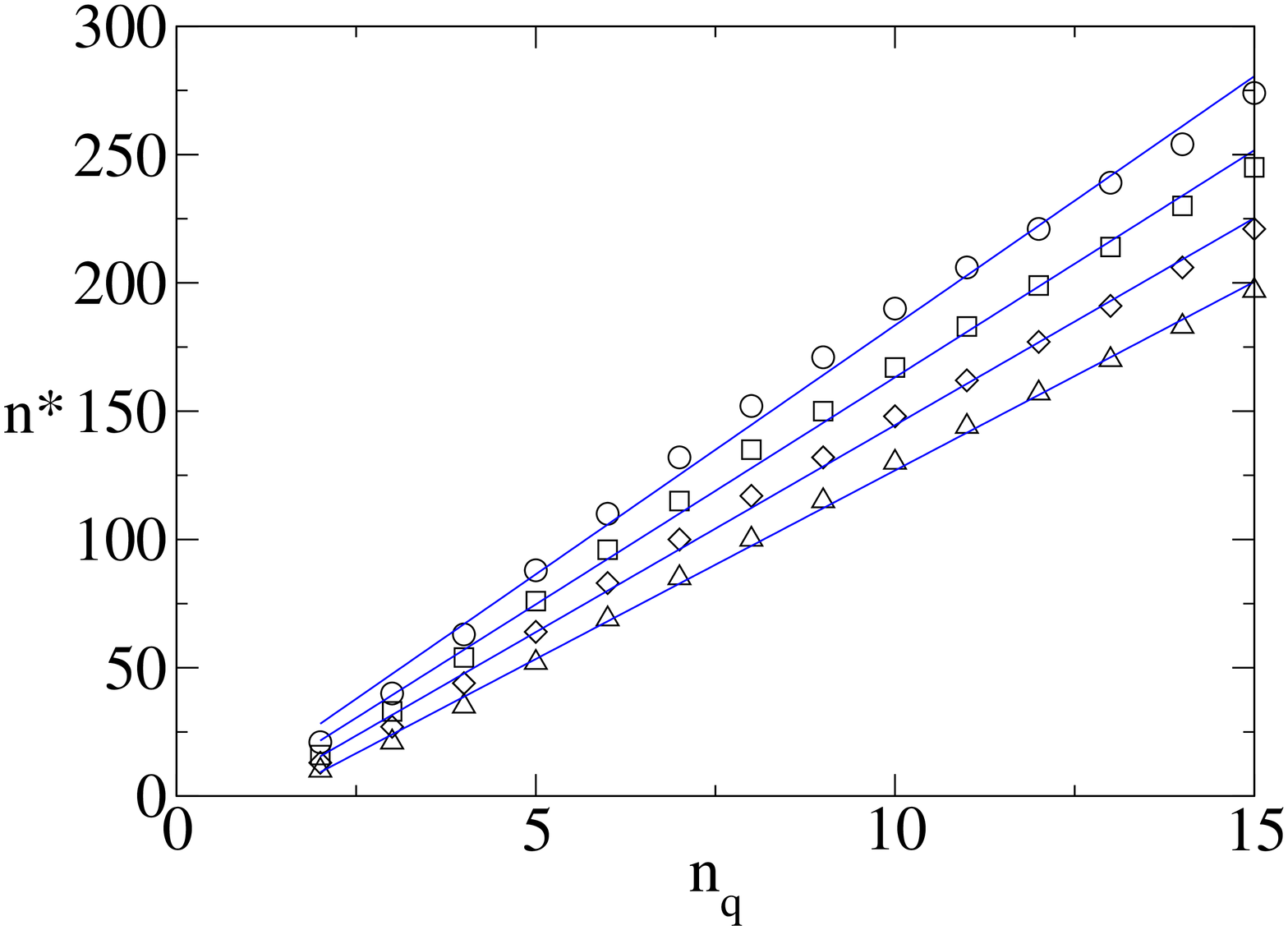,width=4cm,angle=270}
\epsfig{file=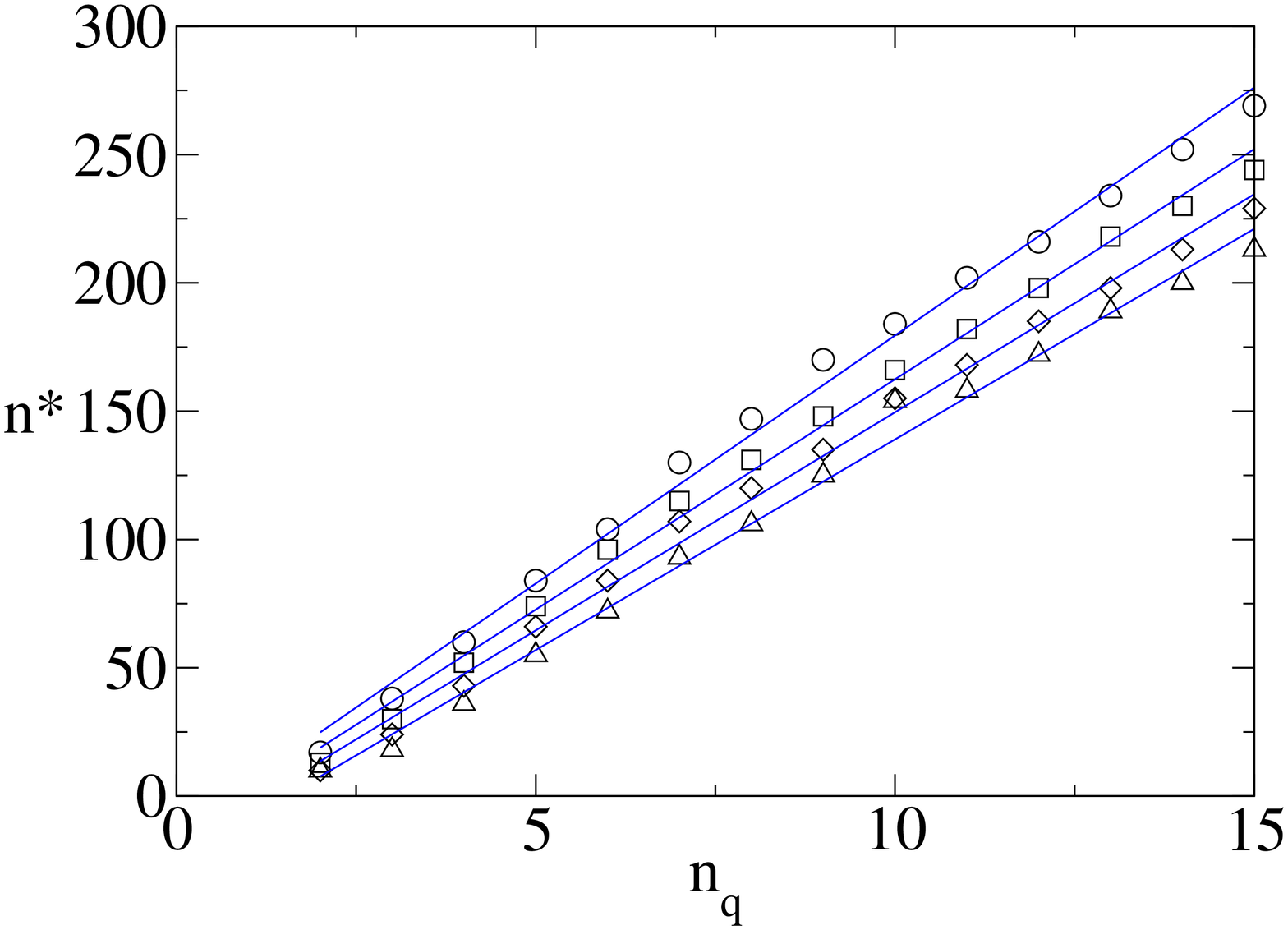,width=4cm,angle=270}
\caption{(Color online) The number of gates $n^*$ needed to achieve
  $D_{\mu_k}\le 
  \epsilon$ as function of the number of qubits together with fits to the
  function $f_1$. Plots 1,  
  2, and 3 correspond to $k=2,4,8$, respectively. 
The different symbols in a plot ($\triangle$,
  $\diamond$, $\Box$, $\circ$) represent different values of $\epsilon$,
  with $\ln(\epsilon)=$0.5, -1.5, -2.5 and -3.5  for $k=2$,
  $\ln(\epsilon)=$-1, -2, -3 and -4 
 for $k=4$, and 
  $\ln(\epsilon)=$1, 0, -1 and -2,  respectively for $k=8$.} 
\label{fig.mom2.result}        
\end{figure}

\begin{figure}[h]
\epsfig{file=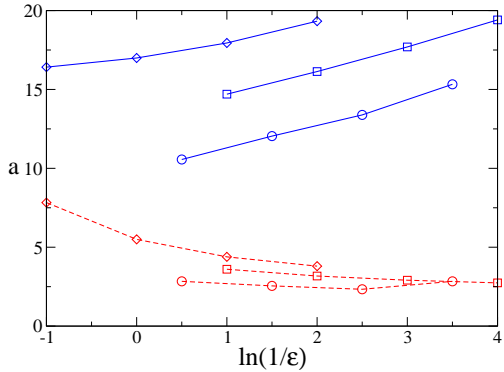,width=6cm,angle=270}
\caption{(Color online) The coefficients $a_1$ (blue full lines) and $a_2$
  (red dashed lines) for the convergence of $D_{\mu_2}$ (circles),
  $D_{\mu_4}$  
(squares), and $D_{\mu_8}$ (diamonds)
  as a function of the available values of $\ln(1/\epsilon)$.} 
\label{fig.mompentelin}        
\end{figure}

In Fig.~\ref{fig.mom2.result}
 we plot the number $n^*$ needed to achieve a fixed small value 
 $D_{\mu_k}<\epsilon$ for all three moments studied, $k=2,4,8$, and for
 different values of $\ln(\epsilon)$, together with fits to the function
 $f_1$ introduced above. The similarity between the three curves is striking.
 The Figure demonstrates that  $n^*(n_q)$  is very well described by a
 linear behavior for all $k$, and that furthermore even the slopes of that
 linear behavior are very similar for all moments.
We have also fitted the data to $f_2$, but our range of $n_q$ is too small to
decide which of the two functions $f_1$ and $f_2$ describes the
scaling with $n_q$ better. In fact, the numerical data $n^*(n_q)$ for
$D_{\mu_4}$ and $D_{\mu_8}$ show a
slight negative curvature, which makes $f_2$ nominally fit worse than
$f_1$. However, 
we believe that the slight negative curvature is a numerical artifact
explained below, and secondly, $f_2$ also represents the dependence on
$\epsilon$ very well. 
This is shown in Figure
 \ref{fig.mompentelin}, where we have collected the coefficients $a_1$ and
 $a_2$ for all 
 moments considered. Figure \ref{fig.mompentelin} shows  that the
 $\epsilon$ dependence is correctly captured by the function $n_q\ln
 (n_q/\epsilon)$: 
 $a_2$ becomes basically independent of $\epsilon$ for small enough
 $\epsilon$. Moreover, the prefactors $a_2$ of all moments
 considered converge to 
 practically the same value once $\epsilon$ is small enough, underlining
 once more the very similar convergence behavior of the three moments with
 $k=2,4,8$.  
The apparent slight sub-linear behavior of
$n^*(n_q)$ for $D_{\mu_4}$ and $D_{\mu_8}$ finds its explanation in the
fact 
that the 
saturation levels of our numerical data for $D_{\mu_4}$ and $D_{\mu_8}$ are
higher than for $D_{\mu_2}$ and $D_P$, such that the possible
values we can choose for $\epsilon$ are closer to saturation than in
$D_{\mu_2}$. This slightly overestimates $n^*(n_q)$, but less so for large
$n_q$, where the approach to saturation is slower, such that the curve
$n^*(n_q)$ appears to curve downwards.

One might wonder if the slight negative curvature may not result from
averaging 
over the column of the matrix. Indeed, while in CUE 
all matrix elements are equivalent in the sense that $\langle
|U_{ij}|^{2k}\rangle$ is independent of $i,j$, and additionally averaging
over a column 
would therefore give exactly the same result, this is not the case in a UCE
circuit of given finite length $n_g$. For example, after one gate, the first
element $U_{11}$, (where the index 1 signifies the state $\ket{0\ldots0}$ in
the computational basis) is never affected by a CNOT, whereas others
are. One effect of the convergence of UCE 
to CUE is that these inhomogeneities decay. One might suspect that averaging
over the first column 
effectively reduces the inhomogeneities and could therefore provide a
mechanism of accelerated convergence compared to a moment that has not been
averaged over a column of $U$. As the sample size (in the sense of the
number of elements in a column used to average) increases exponentially with
$n_q$, small differences in the   $\langle
|U_{ij}|^{2k}\rangle$ are rapidly averaged out, and this might suggest a more
rapid convergence than for a single matrix 
element. Moreover, the effect is expected to become more pronounced
for higher moments, which amplify small initial differences.  

To 
test this hypothesis we calculated for restricted sample sizes ($n_q\le 10$ and
$n_r=10^4$) the forth and eighth moments $\mu_4'$ and $\mu_8'$ for a fixed
matrix  
element (we chose $U_{11}$ and $U_{31}$), defined as in (\ref{muUCE}) but
without averaging over  
the first column. For 
larger values of $n_q$, a 
calculation that does not use averaging over a 
column is unfortunately beyond our numerical capacities. 
The corresponding signals
$D_{\mu_4}'$ and $D_{\mu_8}'$ for the element $U_{11}$ start off at a
  larger value than $D_{\mu_4}$ and $D_{\mu_8}$, and decay more
  rapidly, till the latter are reached. However, this happens at rather
  small
  values of $n_g$, whereas for larger $n_g$, the two curves $D_{\mu_k}'$ and
  $D_{\mu_k}$ for the same $k$ are basically
  indistinguishable within the precision of the data. Thus, averaging over a
  column does not significantly change $n^*(n_q)$.

On the other hand, we verified that also in $D_{\mu_2}$ and in $D_P$ a
slight negative curvature of $n^*(n_q)$ can be produced by pushing
$\epsilon$ close to the saturation level. Furthermore, the
quality of the fits to $f_1$ and $f_2$ deteriorates for decreasing
$\epsilon$. From a physical
perspective a sublinear behavior seems impossible, as it would mean 
that the global state of a large enough quantum circuit equilibrates before 
even every qubit is touched by a quantum gate.  All these elements confirm
the explanation of the slight 
negative curvature as numerical artifact as discussed above.

The main messages from Figs.~\ref{fig.mom2.result} and
Figs.~\ref{fig.mompentelin} is that 1.) all moments considered converge at
basically the same rates; 2.) the number of gates needed to achieve a given
precision increases in good approximation linearly with the number of
qubits, and 3.) the additional $\epsilon$ dependence is well accounted for
by a $n_q\ln(n_q/\epsilon)$ behavior of $n^*(n_q)$.

\subsection{Correlations between matrix elements}
\begin{figure}[h]
\epsfig{file=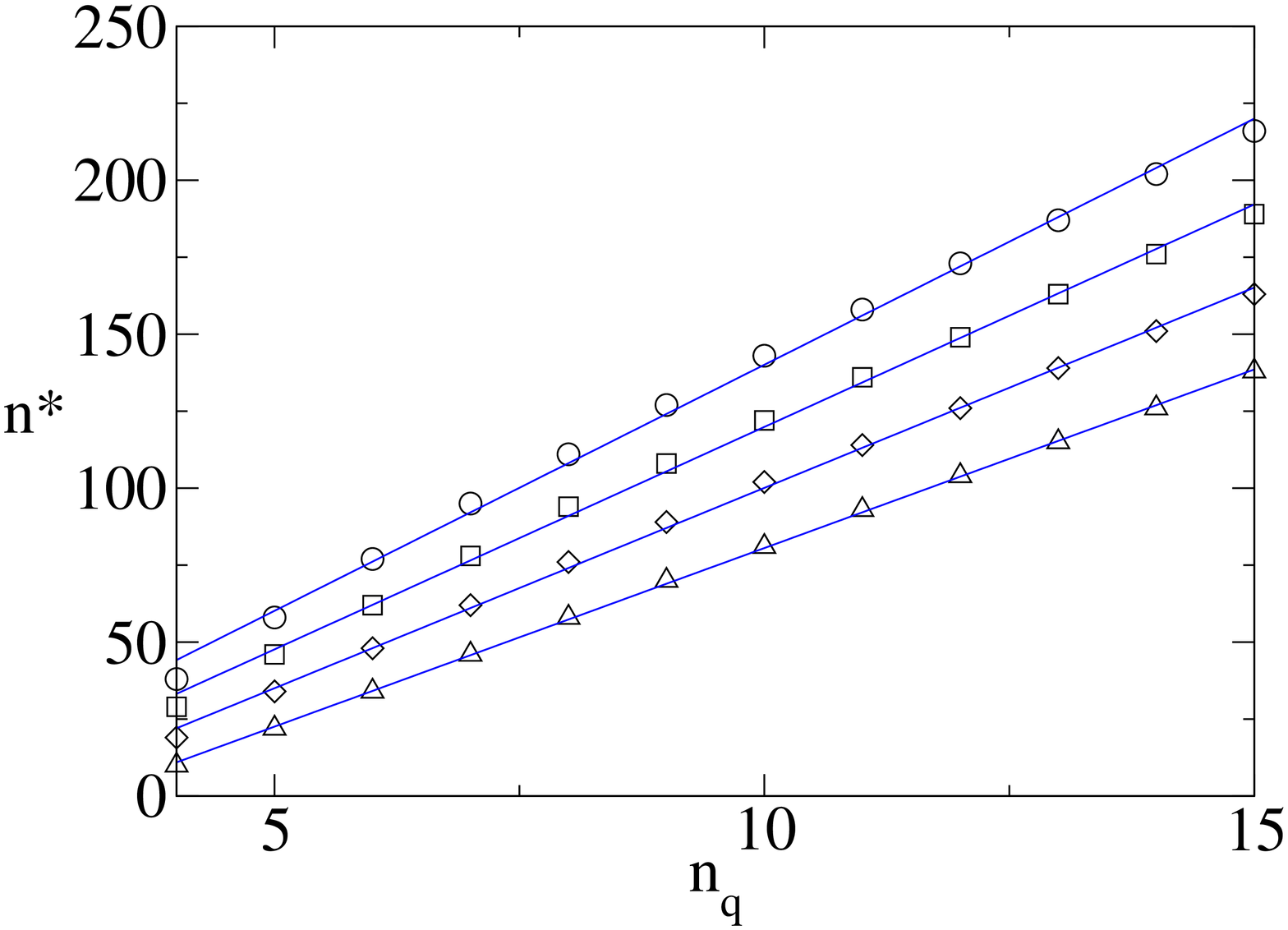,width=4cm,angle=270}
\epsfig{file=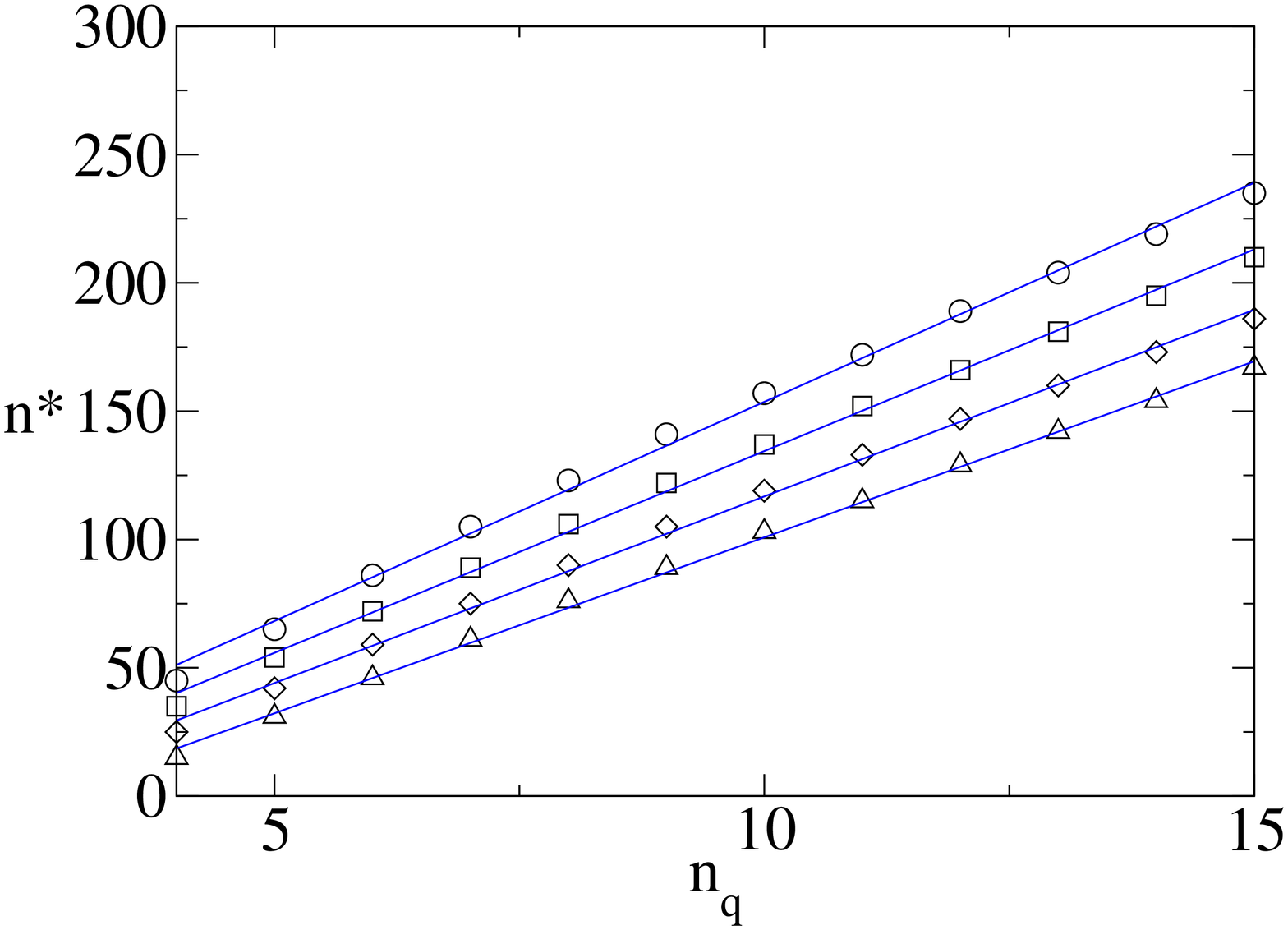,width=4cm,angle=270}
\epsfig{file=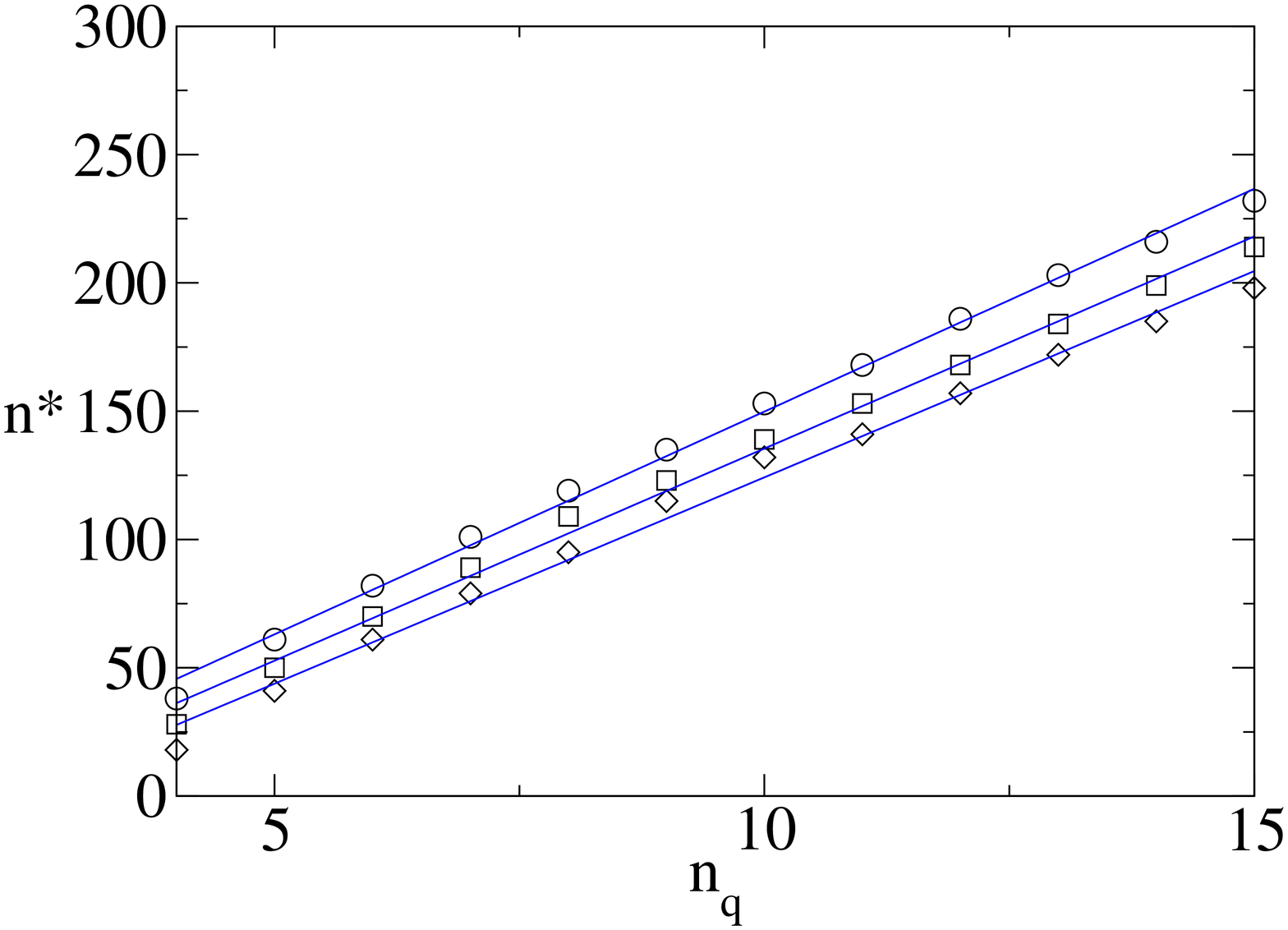,width=4cm,angle=270}
\caption{(Color online) The number of gates $n^*$ needed to achieve
  convergence of the correlators $c_k$, 
  $D_{c_k}\le 
  \epsilon$ for $k=2,4,8$  together with fits to the functions
  $f_1$ (plots 1,2,3, respectively). For $k=2$,
  $\ln(\epsilon)=$-1, -2, -3, and -4 ($\triangle$, $\diamond$, 
  $\Box$, $\circ$); for $k=4$, $\ln(\epsilon)=$0,-1,-2 and -3 ($\triangle$,
  $\diamond$, 
  $\Box$, $\circ$); and for $k=8$, $\ln(\epsilon)=$1, 0 and -1 ($\diamond$,
  $\Box$, $\circ$, 
  respectively).} 
\label{fig.corr2.result}        
\end{figure}

\begin{figure}[h]
\epsfig{file=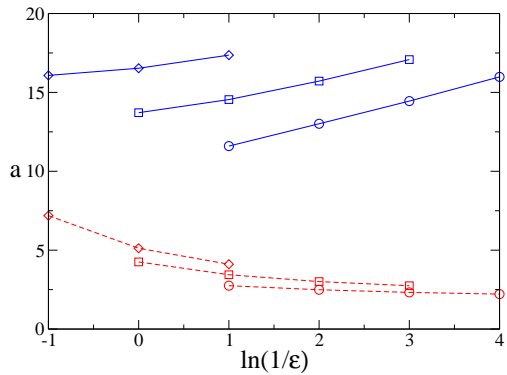,width=6cm,angle=270}
\caption{(Color online) The coefficients $a_1$ (blue full lines) and $a_2$
  (red dashed lines) for the convergence of $D_{c_2}$ (circles), $D_{c_4}$ 
(squares), and $D_{c_8}$ (diamonds)
  as a function of the available values of $\ln(1/\epsilon)$.} 
\label{fig.corpentelin}        
\end{figure}

Even in CUE, different matrix elements are not independently distributed (in
contrast to CUE's Hermitian cousin GUE). One obvious reason for the
correlations is the ortho--normalization of columns and lines of a unitary
matrix. We define correlations between $k$ different $y_{ij}$ for a same
column $j$ as $c_k=\langle\prod_{i=1}^{k}y_{ij}
\rangle=N^k\langle\prod_{i=1}^{k}|U_{ij}|^{2} \rangle$, where the average is
over the considered ensemble. In the CUE case, one finds through invariant
integration \cite{Aubert03}
\be
c_k=\frac{N^{k}(N-1)!}{(N+k-1)!}\,,
\ee 
which differs from $\mu_k$ by a factor $\frac{1}{k!}$. Thus, for small $k$,
the correlations are important, and comparable to the moments of the same
order.  
For UCE we again
average over the column, but include each element in at most one product in
order not to create additional artificial correlations between the products, 
\ben \label{cUCE}
\tilde{c_k}=\frac{1}{n_r[\frac{N}{k}]}\sum_{r=1}^{n_r}
\sum_{i=1}^{[\frac{N}{k}]}\prod_{j=1}^{k}y_{(k\,i-k+j)1}^{(r)}\,.  
\een
Here, $[x]$ means the integer part of $x$. We measure the distance to CUE
as a relative deviation of $\tilde{c_k}$ from the CUE value.  
\bean \label{distance_c}
D_{c_k}&=&\frac{|\tilde{c_k}-c_k|}{c_k}\,.
\eean
Fig.~\ref{fig.corr2.result} shows results for the evolution of $n^*(n_q)$ in
the 
cases $k=2$, $k=4$, and $k=8$, as well as fits to $f_1$. The
behavior is predominantly linear, and very similar for all moments
considered. The numerical data can also be fitted very well to $f_2$, and
again it is difficult within the limited range of $n_q$ values available to
us to clearly 
distinguish between one or the other. The function $f_2$ fits in general
somewhat worse (plot not shown) than $f_1$, but this is due to the same
numerical 
artifact of  slight negative curvatures of
$n^*(n_q)$. Nevertheless, from Fig.~\ref{fig.corpentelin} which
shows the 
fitted coefficients $a_1$ and $a_2$ for all three correlators, it is clear
that the $\epsilon$ dependence is correctly described by $
n_q \ln (n_q/\epsilon)$, and that the prefactor $a_2$ is largely independent
of the order of the correlator for sufficiently small $\epsilon$. 
Moreover, comparing Figs.~\ref{fig.corpentelin} and \ref{fig.mompentelin},
we see that the correlations $c_k$ converge basically with the same rates as 
the moments of the same order, $\mu_k$ --- a result to be expected from the
theory of $k$--designs \cite{Harrow08,Dankert06}. In fact, an alternative
definition of a unitary 
$k$--design is that any polynomial in the complex matrix 
elements of degree $(m,l)$ with $m,l\le k$ has the same average over the
unitary design as over the full unitary group \cite{Dankert06}. 

\section{Conclusion}
We have studied the convergence of the distribution of matrix elements,
moments of that distribution up to $\langle |U_{ik}|^{16}\rangle$, as well
as correlations between matrix elements with up to 16 factors $|U_{ik}|$ 
within the same column, for random quantum algorithms drawn from 
the Unitary Circuit Ensemble (UCE) to their counterparts in CUE. Simulating
quantum circuits with up to 28 qubits (for the distribution of matrix
elements), and up to 18 (15) qubits for the moments (correlations), we have
shown that all these quantities can be efficiently reproduced with a precision
$\epsilon$ using quantum
circuits from UCE containing a number of gates that scales at most as
$n^*\le Cn_q\ln
(n_q/\epsilon)$ with the number of qubits $n_q$, where $C$ is a positive
constant. Such fast convergence comes 
somewhat to a surprise, as for general two--copy gapped gate sets with a gap
$1>\Delta >0$ a quadratic upper bound, $n^*\le
Cn_q(n_q+\ln(1/\epsilon))$, has been shown \cite{Harrow08} (UCE has spectral
gap $\Delta\simeq 0.232703$). While it is
clear that in order to faithfully reproduce the full joint probability
distribution of CUE using UCE circuits one needs a number of gates which scales
exponentially with the number of qubits, our results
suggest that the inefficiently reproduced quantities must be of more
complex nature than the low 
moments of the distribution of absolute values of matrix elements and
their low--order correlation 
functions.

{\em Acknowledgments:} We thank Aram Harrow and Lorenza Viola for useful 
correspondence, G\'eza T\'oth for useful discussions, and CALMIP (Toulouse)
and 
IDRIS (Orsay) for 
the use of their computers. This work was supported by the Agence
National de la Recherche (ANR), project INFOSYSQQ, and the EC
IST-FET project EUROSQIP.
\bibliography{../mybibs_bt}

\end{document}